\newcommand{\bea}{\begin{eqnarray}}
\newcommand{\eea}{\end{eqnarray}}
\documentclass[aps,pra,preprint,amsmath,amssymb,showpacs]{revtex4}
\usepackage{epsfig}
\usepackage{times}
\usepackage{amsmath}
\usepackage{mathrsfs}
\usepackage{graphicx}
\usepackage{epsfig}
\usepackage{dcolumn}
\usepackage{color}
\usepackage{bm}
\usepackage[caption=false]{subfig}
\usepackage{soul}

\begin {document}
\title{Photoassociative  cooling and trapping of a pair of interacting atoms}

\author{Subrata Saha$^1$, Somnath Naskar$^{1,2}$ and Bimalendu Deb$^{1,3}$}
\affiliation{$^1$Department of Materials Science, Indian Association
for the Cultivation of Science (IACS),
Jadavpur, Kolkata 700032, India, \\ $^2$Department of Physics, Jogesh Chandra Chaudhuri College, Kolkata-700033, India, 
$^3$Raman Centre for Atomic, Molecular and Optical Sciences, IACS,
Jadavpur, Kolkata 700032, India.}
\begin{abstract}
We show that it is possible to cool interacting pairs of atoms  by a lin $\perp$ lin Sisyphus-like 
laser cooling scheme using counter-propagating photoassociation (PA) lasers.
 It is shown that  
the center-of-mass  motion (c.m.) of atom pairs can be trapped in molecular spin-dependent periodic potentials generated 
by the lasers. The proposed scheme is most effective for narrow-line PA transitions.
We illustrate this with numerical calculations using fermionic $^{171}$Yb atoms as an example.   
\end{abstract}

\maketitle

\section{Introduction}

The development of laser cooling and trapping \cite{hansch1975, wineland1975, rmp:1998} of atoms over the last 
four decades  has enabled a number of breakthrough achievements such as the realization of  Bose-Einstein condensation \cite{bec:1995},
Fermi degeneracy  \cite{jin:science:1999}, 
Fermi superfluidity \cite{ketterle:nature:2006}, superfluid-Mott transitions \cite{bloch:2003:nature} 
and so on. Illumination of slowly moving atoms with multiple laser beams in a specific geometric configuration allows one to  manipulate
both the internal and external degrees-of-freedom of  atoms.
The internal degrees-of-freedom, such as the electronic configuration or the spin polarization of atoms can be manipulated using circularly polarized resonant 
light as demonstrated by Kastler \cite{Rad:1950} more than 60 years ago. The external degrees-of-freedom such
as the position and  momentum  of atoms can be controlled using radiative forces \cite{Sc:1980,pr:1981,RMP:1986,phystoday1990, 
book:lasercooling:metcalf}. 
Dispersive forces, also called dipole forces arise due to  position-dependent light shifts, leading to  
optically generated  lattice for trapping atoms in an ordered array. The techniques of laser cooling and trapping of atoms developed so far have been essentially 
a single-atom phenomenon, although 
the possibility of a collective laser cooling scheme was theoretically discussed by Vuletic and Chu \cite{prl:2000:chu} about 15 years ago. 

Here, we propose  a  Sisyphus-like laser cooling scheme where interatomic interactions 
play an essential role. Consider the motion of two  colliding atoms. The total kinetic energy of the atom-pair 
consists of relative and the center-of-mass (c.m.) energy   $E$ and $E_{{\rm c.m.} }$, respectively. Cooling 
both atoms simultaneously implies reducing either or both of these kinetic energies. Such cooling by lasers will 
require photoassociative transitions \cite{prl:1987:julienne,rmp:1999:weiner,rmp:2006:jones} that are cyclic. In 
photoassociation (PA), a single photon connects the state of two interacting ground-state atoms  with an excited molecular 
bound state via so-called free-bound dipole transitions. For PA to occur, relatively cold atoms with temperatures typically at  
or below millikelvin regime are required. Usually, the c.m. motion of atom-pairs is not of much spectroscopic interest and hence neglected in PA spectroscopy.
However, at a fundamental level, both 
relative and c.m. motion of atom-pairs become coupled by PA process. The question then naturally arises as to whether it is possible to reduce the kinetic 
energies of the two coupled-motion  from millikelvin regime to lower temperatures by Sisyphus mechanism 
using photoassociative transitions. We show that it is indeed possible by coherent manipulation of   
the coupled motion of atom-pairs with counter-propagating PA lasers in lin $\perp$ lin configuration. Coherent 
coupling in a PA system requires that the excited molecular states to which PA lasers 
are tuned should have life time long and a relatively large Franck-Condon (FC) overlap integral with the scattering state of ground-state atoms \cite{prl:2013:yan,prl:2016:taie}. The recent experimental demonstrations of Rabi oscillations in two-electron atomic 
systems such as Sr \cite{prl:2013:yan} and Yb \cite{prl:2016:taie} making use of ultra narrow 
intercombination  PA transitions strongly suggest that such PA systems can be treated in a way somewhat analogous to the methods used for two-level atoms.

Sisyphus cooling of atoms, first proposed by Pritchard \cite{prl:1983:pritchard}, 
 involves an atom having degenerate sub-levels in its electronic ground and excited states. It uses 
counter-propagating polarized  lasers to produce spin-dependent and spatially modulated light shifts  that eventually lead to sub-Doppler  cooling. 
We here propose a photoassociative Sisyphus-like  method with  sub-levels corresponding to molecular angular momenta. 
For two ground-state 
atoms in collision, 
the  degenerate sub-levels correspond to multiple scattering channels having the same asymptotic threshold.  
A PA system can  access a single vibrational and rotational level in an electronically excited molecular potential. 
In narrow-line PA systems, cyclic PA transitions or Rabi oscillations  between 
an excited molecular bound state and the continuum of ground-state scattering states are possible as can be evidenced from experiments \cite{prl:2013:yan,prl:2016:taie}.  

For single atoms, Sisyphus cooling relies on optical pumping in  multilevel atoms moving in an optical field with  spatially varying polarization.   
 Optical pumping can also be applicable for continuum-bound as well as continuum-continuum transitions in diatomic molecules 
 as discussed by  Mies \cite{fhmies}  about 
 35 years ago. The ``continuum''  here refers to the dissociation continuum of  scattering states between two free atoms. In case of 
 PA, the initial motional state of two atoms is a scattering state which can be considered as the dissociation continuum
 of the ground-state molecule. Dark state resonances and optical pumping into an atom-molecule dark state in two-color PA of ultracold atoms have been 
 experimentally observed \cite{prl95:2005:winkler,prl96:2006:moal}. The physical interpretation of dark state resonance and optical pumping 
 in two-photon PA of ultracold atoms has been discussed by Cohen-Tannoudji \cite{physscripta:2015:cohen-tannoudji}.  
 Another coherent effect related to dark state 
 resonance is stimulated Raman adiabatic passage (STIRAP) which has been extensively investigated by Bergman and coworkers \cite{rmp:1998:bergman}
 in three level atomic systems. STIRAP with a continuum as an initially populated or 
intermediate or a final target state has also been studied by many workers
\cite{Opt:1992,prl:2005:halfmann,prl:1992:hioe,pra76:2007,CHEM:2012} over the years.  
In recent times, STIRAP using PA of ultracold atoms has been discussed and debated by many authors 
\cite{pra60:1999,pra58:1998,pra65:2002, pra65:2002:java,pra64:2001,prl:2000:java}. 

The elementary process underlying our method 
is schematically shown in Fig.{\ref{fig:fig1}}.  Let a pair of slowly moving atoms  
be  subjected to a pair of lin $\perp$ lin  counter-propagating PA laser beams having the same frequency.  
 Let us consider, for the sake of simplicity, two colliding 
ground-state atoms 1 and 2 with 
total energy $E_t = E + E_{{\rm c.m.}}$ and total angular momentum $J_g=1$ are being acted upon by the   
lin$\perp$lin  counter-propagating PA lasers. Suppose, the lasers are tuned near resonance to a particular  excited bound state  $ |b \rangle \equiv |v, J_e \rangle$, 
with  vibrational  quantum number $v$ and rotational quantum number $J_e$ from a relative energy having significant free-bound FC factor.  
We here closely follow the Sisyphus method of Dalibard and Cohen-Tannoudji \cite{cohen-tannoudji:josab},
but generalize for  photoassociative transitions.  The physical processes underlying our proposed photoassociative cooling of atom-pairs 
differ from those of standard Sisyphus method of cooling of single atoms in the following respects: First, in case of standard Sisyphus method there 
is only one kind of external degrees of freedom of motion which is the c.m. momentum of single atoms while in our proposed 
photoassociative Sisyphus method,  two  kinds of external degrees-of-freedom of motion are involved: these are the relative momentum 
${\mathbf p}$ and the c.m. momentum ${\mathbf P}_{\rm c.m.}$ of the two colliding atoms.

In an elementary process of PA where one photon from PA laser 
is absorbed followed by spontaneous emission of one photon from the excited molecular state, the momentum conservation dictates that the change in c.m. 
momentum $\Delta {\mathbf P}_{{\rm c.m.}} =  {\mathbf P}_{{\rm c.m.}}^{{\rm final}} - {\mathbf P}_{{\rm c.m.}}^{{\rm initial}}$, where 
${\mathbf P}_{{\rm c.m.}}^{{\rm initial(final)}}$ denotes the initial(final) c.m. momentum, should satisfy 
\bea 
\Delta {\mathbf P}_{{\rm c.m.}} =  \hbar \left [ {\mathbf k}_{{\rm PA}} - {\mathbf k}_{{\rm spon}} \right ] 
\eea 
with ${\mathbf k}_{{\rm PA}}$ and ${\mathbf k}_{{\rm spon}}$ being the momentum of PA laser photon and spontaneously emitted photon, respectively.
Since the relative
motion of the two atoms occurs under the influence of molecular potentials, the relative momentum 
 ${\mathbf p}$ is not a good quantum number to specify  the conservation of momentum in this situation. 
  Second, in case of standard Sisyphus method, energy exchange between the atoms and photons occurs 
 at the expense of the c.m. motional energy of single atoms, but in our proposed scheme  energy exchange can happen 
 between photons and the total motional energy $E_t$ which is a sum of $E = p^2/2\mu$ and $E_{{\rm c.m.}} = P^2/2M$ where 
 $\mu$ and $M$ denote the reduced and total mass of the atom-pair. This means that energy transfer can happen to 
 both the c.m. and relative motions of the atom-pair from the photonic fields.  
 This is because PA process couples both relative and c.m. motion. 
 Recently, coupling between the relative and c.m. motion  of exactly two atoms in an anharmonic trap has been experimentally demonstrated by Sala {\it et. al.} \cite{sala:prl110}. These experimental results and the recent observation of 
 Rabi oscillations in PA \cite{prl:2013:yan,prl:2016:taie} indicate that a two-level type treatment of a free-bound system is possible.
 Third, by standard Sisyphus method one can trap single atoms  while 
 by our proposed method one can trap the c.m. motion of  atom-pairs. In this context, it is to be noted 
that, unless PA occurs in a tightly confined trap,  the trapping potential has practically no influence over PA transitions since PA 
 occurs near the trap center where trapping potential is negligible.  
 The interatomic separations at which PA transitions take place are typically far below or  of the order of a nanometer. So, unless 
 trapping size is reduced to a nanometer or less, PA process will not be affected by the trapping potential \cite{prl:2016:taie}. So, to subject a pair of cold atoms into counter-propagating PA lasers, one can prepare pairs of atoms in a two-atom Mott insulator where each lattice site is occupied by two atoms as is done in the recent experiment by Taie {\it et. al.} \cite{prl:2016:taie}. PA lasers will then act almost independently 
on separated atom-pairs in optical lattice. In that case, the lasers that generate the lattice should be far off PA resonance so that they do not disturb the PA process. Once the atom-pairs are cooled enough for their c.m. motion to be trapped by dipole forces, the 
lattice lasers may be switched off since the c.m. motion of the pairs is trapped in the periodic dipole potentials  generated by PA lasers.

 In contrast to Sisyphus cooling of single atoms where only one kind of external motional energy scale is involved, in the present case there are two external energy scales: c.m. and relative motional energy. The change in c.m. momentum of the atom-pair is governed by the principle of momentum conservation of light scattering. The  energy conservation is maintained by a decrease in total energy $E_t$ when the spontaneously emitted photon carries away the excess energy that is released from the coupled relative and c.m. motion. Two atoms colliding with opposite momenta will have maximum probability to come closer to each other and so to couple with PA laser. This means that an atom-pair with zero or very small c.m. energy is most likely to be influenced by PA laser.
We assume that the c.m. kinetic energy $E_{{\rm c.m.}}$ is much 
smaller than the relative motional energy $E$,  PA resonance is then primarily determined by $E$ when the lasers are tuned close to an excited molecular bound level from the threshold of ground-state continuum. The PA detuning parameter $\delta_E = E/\hbar + \omega_L - 
\omega_b$ is an explicit function of $E$, where $\hbar \omega_b$ is the bound-state energy measured from the continuum threshold and $\omega_L$ is the angular frequency of the PA laser.

    The strength of PA coupling is given by Frank-Condon (FC) overlap factor. In the low energy regime, the FC factor as a function of $E$ shows a prominent 
peak at a particular energy $E=\bar{E}$. Let  $\delta_{E=\bar{E}} = 0$ if $\omega_L = \bar{\omega}_L$. So, $\bar{E}$ and $\bar{\omega}_L$ can be termed as resonance energy corresponding resonance laser frequency $\bar{\omega}_L$. In the weak-coupling limit, Sisyphus cooling remains effective so long as the laser detuning  $|\delta| \le \gamma$ or  $|\delta| \simeq \gamma$, where $\gamma$ is the spontaneous linewidth. So, the spontaneous emission into the continuum from the excited bound state may lead to continuum states with an energy spread of the order of $\hbar \gamma$. If the final energy $E_f$ after the spontaneous emission is less than $\bar{E}$ we have cooling effect.

 On the other hand, if $E_f$ is greater than $\bar{E}$, the atoms may go out of cooling cycle. 
Therefore,  we need a repumping mechanism to bring back the atoms that acquire higher relative energy as a result of spontaneous emission. This can be accomplished by applying  counter-propagating broadband PA lasers in $\sigma_+$ + $\sigma_-$ configuration, in addition to the monochromatic Sisyphus cooling lasers. These broadband repumping lasers should have
the same central frequency $\bar{\omega}_L - \gamma$ such that the  pair of atoms having energy $E \simeq \bar{E} + \hbar \gamma$ 
comes to PA resonance with the central frequency and  are re-pumped into the cooling cycle.  We consider $\sigma_+$ + $\sigma_-$ configuration for repumping lasers, that is,  one laser with $\sigma_+$ 
and the other $\sigma_-$ polarization, because this configuration leads to space-independent light shift of the Zeeman sub-levels, redistributing the population in 
both ground- and excited- state sub-levels \cite{cohen-tannoudji:leshouches:1992}. In the present context, the Zeeman sub-levels refer to the total magnetic quantum number of 
the two atoms because the two atoms are correlated and the photoassociative  transitions occur between the correlated two-atom state  and the molecular bound state. 

  Since the repumping lasers give a space-independent shift for all the sub-levels, there is no additional optical force resulting from this light shift. Regarding the bandwidth of the repumping lasers, the bandwidth may be set at a few times of $\gamma$ or one order of magnitude larger than $\gamma$. If $\gamma$ is in the kHz regime as 
in the case of metastable excited states or spin-forbidden intercombination transitions in Yb and Sr atoms, the bandwidth of the repumping lasers would be in MHz regime. For larger $\gamma$, 
the bandwidth of the repumping lasers has to be large enough. The repeating cycles of the repumping laser pulses should be much smaller than $\gamma$ so that 
once the atom-pair is excited to the bound state, the pair can quickly decay back to the ground-state continuum without any memory effect of the pulse \cite{josab:27:2006:monroe}.

Broadband laser cooling \cite{book:laser-light-pressure:letokhov:1987,OptLett:13:Hoffnagle}  with 
 $\sigma_+$ + $\sigma_-$ scheme is theoretically described by Parkins and Zoller \cite{pra:45:1992}. Doppler cooling of atoms or atomic ions with broadband or pulsed lasers has been experimentally demonstrated  \cite{prl:67:1991:zhu,josab:27:2006:monroe}. In recent times, broadband laser cooling methods have been successfully applied 
 for vibrational or rotational cooling of some specific molecules \cite{naturecomm:2014,science:pillet,PRA:80:2009:Sofikitis,PRL:109:2012}. 
 Our proposed scheme may be compared to Sisyphus laser cooling of diatomic molecules  \cite{science:pillet,pra:2014:comparat,molphys:manai}. 
 The c.m. and relative motion of atom-pairs are equivalent to translational and vibrational motion of a diatomic molecule. 
 In general, laser cooling of molecules
 is extremely difficult due to the presence of a plethora of ro-vibrational levels hindering cyclic transitions that are needed for laser cooling. 
 Nevertheless,  
 direct laser cooling of some specific molecules \cite{science:pillet,molecule-cooling} has been successfully demonstrated in recent times.
 
  The optical pumping has been used to demonstrate vibrational cooling  of molecules  in a recent experiment \cite{science:pillet}. 
 These recent advances towards laser cooling of diatomic molecules and coherent PA \cite{prl:2013:yan,prl:2016:taie} 
 motivate us to explore Sisyphus-like scheme to cool and trap an analogous system of a pair of atoms in a state of slow collision. This will 
 allow one to optically control both the relative and c.m. motions between two atoms.

In our proposed scheme, we have  two coupled and competing variables involved - relative and c.m. momentum. The relative momentum is a fast variable and the c.m. momentum is a slow one.
For narrow PA linewidth, one can thus adiabatically eliminate relative motion leaving c.m. motion to follow 
an optical potential obtained by averaging  over  $E$ near   $\bar{E}$. Since the c.m. energy $E_{\rm c.m.}$ is much smaller than the depth of the optical potential, 
the c.m. motion of the pair will eventually be trapped.
Since the optical pumping will preferentially bring the atom-pair at the bottom of the potential well with the release of mostly relative kinetic energy, by successive absorption-emission cycles the pair will be cooled. In case of single atoms, 
 Sisyphus  or other sub-Doppler cooling and trapping methods  such as velocity selective coherent population trapping (VSCPT) make use 
 of a number of optical coherent effects such as  Rabi oscillations, dressed states, optical pumping, saturation effects, light shifts, 
 dark-state resonances, etc., at the 
 backdrop of randomness introduced by spontaneous emission \cite{cohen-tannoudji:leshouches:1992}. As we embark to extend Sisyphus method to 
 photoassociative continuum-bound systems,
 the question naturally arises as to whether similar coherent effects can be obtained using continuum-bound optical transitions.  
The problem of coherent coupling, Rabi oscillations, and dressed states in continuum-bound coupled systems had been addressed by 
a number of workers  in the
1980s   \cite{prl47:1981:eberly,jphysb:1982:coleman,physscripta:Javanainen} and early 1990s \cite{physreports:1990:knight}, 
particularly in the  context of autoionization or photoionization. It was theoretically 
shown that Rabi oscillations in a continuum-bound system are possible provided the continuum-bound 
matrix element is strong  and the continuum has a narrow resonance \cite{physscripta:Javanainen}. In fact, fulfillment of exactly such conditions have enabled experimental demonstrations of Rabi oscillations in PA \cite{prl:2013:yan,prl:2016:taie}.
Saturation effects in PA have also been studied experimentally by many groups \cite{pra66:Schloder,pra71:kraft,prl91:hullet}.

A class of excited molecular bound states known as purely long-range (PLR)  states \cite{prl:1978}  will  play a particularly 
useful role in our proposed method. These states are localized at nanometer scale separations at which electronic charge overlap of the two atoms may 
be negligible. As a 
result, the number of ro-vibrational levels or scattering channels that can be optically coupled to PLR states is drastically 
reduced. Furthermore, 
in some cases, PLR  states  can  be  accessible by   
 photoassociative transitions from a single scattering channel  \cite{giant,eusophys,freqshift} only. PLR potentials are usually 
 quite shallow, and so capable of supporting only a small number of vibrational levels. As a consequence, the free-bound FC factor at ultralow collision energy can be large. 
 At ultralow temperatures, the 
 collision energy of the initially two free atoms is quite low. This can lead to a narrow PA resonance, 
 providing a unique advantage for generating coherent photoassociative coupling.

  The remainder of the paper is organized in the following way. 
  In the next section,  we present theory of generating 
  molecular spin-dependent periodic optical potentials for cooling and trapping an atom-pair. 
  In Sec.III we discuss  the implementation of our proposed scheme and present numerical results 
  using  a pair of fermionic $^{171}$Yb atoms and their photoassociative coupling to a PLR state. The paper is concluded in  Sec.IV.
  
 \section{The theoretical method}

We now present general mathematical formalism of a simple one-dimensional scheme of photoassociative cooling and trapping (PACT)
as schematically depicted in Fig.1.  In the center-of-mass (c.m.) frame, we can express the Hamiltonian of our system in terms of 
c.m. and relative coordinates  ${\mathbf R} = ({\mathbf r}_1 + {\mathbf r}_2 )/2$ and ${\mathbf r} = {\mathbf r}_1 - {\mathbf r}_2$, 
respectively. The Hamiltonian describing the interaction of a pair of cold atoms with the PA lasers
is $\hat{H}=\hat{H}_0^{{\rm c.m.}} + \hat{H_0}^{{\rm rel}} + \hat{H_I}$; where  
\begin{eqnarray}
 \hat{H_0}^{{\rm rel}} 
=\int_0^\infty E^{'}|E^{'}\rangle\langle E^{'}|dE^{'}+\hbar \omega_b|b\rangle\langle b|
\label{hrel}
\end{eqnarray}
 describes relative motion between the 
two atoms with $|E\rangle $ representing the continuum of ground-state scattering states in the relative 
energy ($E'$) basis and    
$ |b\rangle $ being an excited  molecular bound state. Here 
\begin{eqnarray}
\hat{H}_0^{{\rm c.m.}} = \hat{P}_{\rm c.m.}^2/2M = - [\hbar^2/2M] \nabla_R^2 
\end{eqnarray}
is the kinetic term 
of c.m. motion.  The interaction part   
\begin{eqnarray}
\hat{H_I}=\left[\int_0^\infty\Lambda_{bE^{'}} e^{-i\omega_{L} t} |b\rangle \langle E^{'}|dE^{'}+H.c.\right]
\end{eqnarray}
where  
$\Lambda_{bE} = - \Big\langle b\Big|\mathbf{d_1.{\cal{E}}({\mathbf r}_1)+d_2.{\cal{E}}({\mathbf r}_2)}\Big|E\Big\rangle$ 
is molecular dipole coupling, 
$\mathbf{d_1}$ and $\mathbf{d_2}$ are atomic dipole moments with $\mathbf{r_1}$ and $\mathbf{r_2}$ being the position vectors of atoms 1 and 2, 
respectively.
We can then write $\Lambda_{bE}$ in terms of molecular dipole operator $\mathbf{D}({\mathbf r})$ in the form
\begin{eqnarray}
   \Lambda_{ bE}=- \langle b|\mathbf{D} \cdot \hat{\epsilon}({\mathbf R}) \sqrt{2}\mathbf{{\cal{E}}_0}\cos({\mathbf k}_L 
   \cdot {\mathbf r}/2) |E\rangle
 \end{eqnarray} 
 where $\hat{\epsilon}({\mathbf R})=\left[\mathbf{\hat\sigma}_-\cos({\mathbf k}_L \cdot {\mathbf R} )-i\mathbf{\hat\sigma}_+
 \sin({\mathbf k}_L \cdot {\mathbf R})\right] $ is the polarization vector. Since $\hat{H}_I$ depends both on relative and c.m. coordinates, 
 the relative and c.m. motion becomes coupled. The optically generated force is given by
     \begin{eqnarray}
    \mathbf{ F}&=& \nabla_R \int_0^\infty \Big[\rho_{E'b} \left(\Lambda_{bE^{'}}(R)e^{-i\phi(r,R)} \right)e^{-i\omega_{L} t} +c.c. \Big]dE' 
     \label{for}
     \end{eqnarray}  
 where $\rho_{Eb}$ is the density matrix element representing continuum-bound coherence and $\phi(r,R)$ is the position dependent phase part of the applied laser field. This force $\mathbf{ F}$ is a sum of two forces \textemdash  one is dipole force ${\mathbf{ F}}_{dip}$ and another is dissipative force ${\mathbf{ F}}_{dis}$. 
 
 Here, we consider the simplest one-dimensional 1D model of Sisyphus cooling. A pair of PA  lasers are applied along the $z$-axis, so we have 
 ${\mathbf k}_L \cdot {\mathbf R} = k_L Z $. At $Z = n\lambda/4$ ($n$ is an integer) the polarization is $\sigma_-$ for even $n$ and $\sigma_+$ for odd $n$. 
 After having done a lengthy calculation (see Appendixes A and B), we derive the expression for dipole force as given by
 \begin{figure}[h!]
                      \includegraphics[width=0.45\textwidth,height=0.50\textwidth]{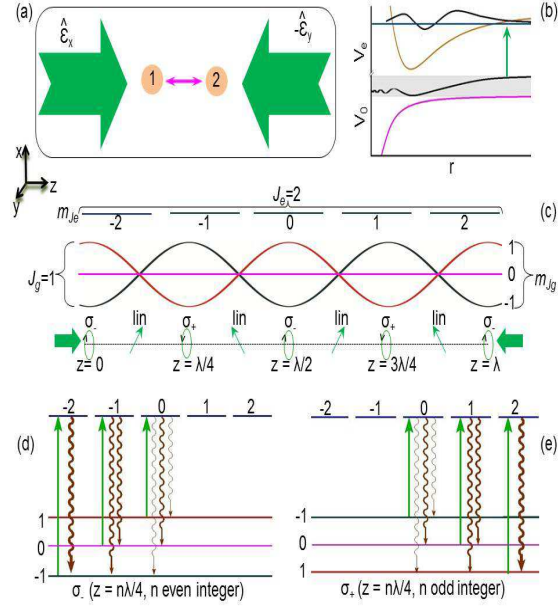}
                     \caption{(a) A  pair of slowly moving atoms 1 and 2  is subjected to a pair of lin$\perp$lin counter-propagating PA lasers 
                      along  $z$-axis. 
                     (b) PA process:  $V_g$ is molecular ground-state potential and $V_e$ is excited PLR potential.
                    The ground-sate scattering continuum state is coupled to a ro-vibrational state in the excited potential by the two lasers.
                    (c) Energy level diagram showing formation of spatially varying (w.r.t. c.m. co-ordinate ($Z$))
                     spin-dependent ground state potentials under the action of applied lasers.
                     (d) Optical pumping (green lines) and spontaneous transition 
                     (curvy brown lines having width proportional to the probability of transition)
                     at mentioned c.m. co-ordinate where light is $\sigma_{-}$ polarized. (e) Same as (d) where light is $\sigma_{+}$ polarized. 
                     An additional pair of counter-propagating broad-band PA lasers in $\sigma_+ + \sigma_{-}$ configuration along the 
                     $z$-axis may be used (not shown in the figure) as repumping lasers (see the text).}  
                     \label{fig:fig1}
          \end{figure}  
 \begin{eqnarray}
    {\mathbf{ F}}_{dip}(Z)=-\frac{\hbar\delta_{\bar{E}}}{\pi}\Big[\frac{\frac{d}{dZ}\Big({\int \limits_{0}^{\infty}}{\cal G}_{E}(Z)d\omega_E\Big)}{\delta_{\bar{E}}^2+\frac{\gamma^2}{4}+\frac{2}{2\pi}{\int \limits_{0}^{\infty}}{\cal G}_{E}(Z)d\omega_E}\Big]
  \end{eqnarray}
 where $\delta_E=\left(\omega_L-\left(\omega_b-\frac{E}{\hbar}\right)\right)$ and $\gamma$ is the  spontaneous linewidth of the bound state. Here the spontaneous emission is taken into account by a master equation approach as elaborated in Appendix B. It is worth mentioning that the master equation approach to discrete systems such as two-level atoms or a single-mode cavity field is well known. However, the master equation approach to a coupled continuum-bound system is not adequately addressed in the literature. Unlike discrete systems, master equation formalism for continuum-bound systems leads to integro-differential equations which are in general difficult to solve analytically. 
 
 As derived in the Appendix A, the optical potential can be expressed as
 \begin{eqnarray}
  U_{opt}&=&\hbar\delta_{\bar{E}}\ln \Big(1+\frac{\frac{1}{\pi}{\int \limits_{0}^{\infty}}{\cal G}_{E}(Z)d\omega_E}{\delta_{\bar{E}}^2+\frac{\gamma^2}{4}}\Big)
 \end{eqnarray}
 In weak intensity limit (as mentioned in appendix A)
  \begin{eqnarray}
   U_{opt}=\hbar\delta_{\bar{E}}\frac{\frac{1}{\pi}{\int \limits_{0}^{\infty}}{\cal G}_{E}(Z)d\omega_E}{\delta_{\bar{E}}^2+\frac{\gamma^2}{4}}
  \end{eqnarray}
 Separating out ${\cal{G}}_E(Z) $ in the form  ${\cal{G}}_E(Z)=\Gamma(E)\Theta(Z)$, we obtain  
 $U_{opt}= \hbar \delta_{\bar{E}} S_0 \Theta(Z)$,
 where
 \begin{equation}
 S_0=\frac{\frac{1}{\pi}{\int \limits_{0}^{\infty}}\Gamma(E)d\omega_E}{\delta_{\bar{E}}^2+\frac{\gamma^2}{4}}
 \label{sat}
 \end{equation} 
is a parameter that describes saturation effect in PA.  In terms of $S_0$ the excited bound state population $\rho_{bb}$ in steady-state can be written as
  
  \begin{eqnarray}
   \rho_{bb}=\frac {S_0 \Theta(R)}{2+2S_0 \Theta(R)}
   \label{rhobb}
  \end{eqnarray}
 $S_0$ can be considered as the photoassociative  counterpart of saturation parameter of a two-level atom. The saturation intensity in TLA is the intensity at which  the parameter becomes unity at resonance. Eq.(\ref{sat}) shows that for $\delta_{\bar{E}}=0$,  $S_0 = 1$ when $ (4/\pi \hbar) \int_0^{\infty} \Gamma(E) d E = \gamma^2$. Note that $\Gamma(E)$ is  proportional to laser intensity. It is now clear that the saturation intensity in PA will depend on FC factor: the larger the FC factor, the lower is the saturation intensity. The amount of FC factor depends primarily on the amplitude of  the scattering wave function at internuclear separations near the outer or inner turning point of the excited  bound state, rather than the scattering phase shift. In this context, an interesting question is the effect of  unitarity regime of scattering on the saturation. The unitarity happens when the scattering phase shift is  $\pi/2$ and as a consequence the magnitude of scattering $T$-matrix element attains its maximum 
value of  
unity. As the phase shift goes through $\pi/2$ as a function of energy, scattering cross section shows a resonance
 structure. Whether this unitarity-limited scattering will maximize FC factor depends on whether the most prominent  anti-node of the scattering wave function at unitarity appears  near the outer turning point of the excited
 molecular bound state. For instance, in case of cold collision between bosonic atoms or between two-component fermionic atoms, the scattering is predominantly of $s$ wave and so the unitarity-limited scattering wave
 function $\psi_0(r)$ asymptotically behaves as $\psi_0(r) \sim \sin(k r + \delta_0) \sim \cos(k r) $, where we have  put the value of the $s$-wave phase shift $\delta_0 = \pi/2$. So, in the limit $k \rightarrow 0$, the scattering wave  function will maximize at asymptotically large separations. Now, on the question as to whether this will lead to the maximization of FC factor obviously depend on the location of outer turning point. If the outer turning point  lies at such large separations where the scattering wave function attains its asymptotic behavior, then the FC  factor will  maximize for a fixed laser intensity. On the other hand, for $p$-wave scattering, the reverse effect  will happen since the $p$-wave scattering wave function $\psi_{\ell=1}(r)$ behaves as  $\psi_{\ell=1}(r) \sim \sin(k r) $ at large separations. The behavior of scattering wave function at short separations can not be analytically predicted {\it a priori}, and  hence no definite relationship of unitarity  regime with saturation effect in PA 
can be  established. One can notice from Eq.(\ref{hrel}) that for large saturation parameter, the population of the excited bound states  $\rho_{bb}$ approaches $\frac{1}{2}$. This means that the other half of the total population  is contained in the ground continuum states when the saturation parameter  is large.
  However, we work much below the saturation limit, i.e.,  $S_0 << 1$.

 Let  $J_e=2$ as in Fig.{\ref{fig:fig1}}. The form of $\Theta(Z)$ is then given by 
\begin{eqnarray}    
\Theta(Z)_{m_{J_g}=\pm1}&=&-A\pm B\cos(2k_LZ)\nonumber\\
\Theta(Z)_{m_{J_g}=0}&=&\text{const}
\end{eqnarray}
where $A$ and $B$ are constants that depend on Clebsch-Gordon(CG) coefficients. We thus obtain light-shifted 
periodic potentials for c.m. motion for different magnetic quantum numbers, that is, we get spin-dependent potentials
for c.m.motion. The c.m. momentum changes due to absorption of a photon  from one plane wave and 
its subsequent emission into another plane wave.  Let us consider a point $ Z=\lambda/4$ where the field is 
 circularly polarized in positive sense. For such polarization, three possible transition pathways may arise Fig.{\ref{fig:fig1}}(e): (1)\hspace*{0.2cm}The continuum 
 state with  at $m_{J_g}=1$ can couple to bound state $m_{J_e}=2$. This state can decay to $m_{J_g}=1$ and all other transitions are forbidden, (2) 
 The  state with $m_{J_g}=-1$ can be excited to $m_{J_e}=0$ state and it can decay to $m_{J_g}=1$ or $m_{J_g}=-1$ or $m_{J_g}=0$ state. (3) The state with  $m_{J_g}=0$ can be excited to $m_{J_e}=1$ state and it can decay to state $m_{J_g}=1$ or $m_{J_g}=0$ 
 state. All the spontaneous transition lines shown here have widths proportional to the square of the corresponding CG coefficients. 
 The net effect is the maximum occupation  probability happens for the state $m_{J_g}=1$ having the lowest c.m. energy at the potential minima.
This argument applies similarly
for locations where the field is circularly polarized in the negative sense as shown in Fig.{\ref{fig:fig1}}(d). 
  Now,  as an atom-pair  moves up the hill from lower potential energy side, it loses its c.m. kinetic energy. After reaching the top of 
 the hill and just about to start gliding down, the optical pumping intervenes,  
 transferring  the pair into a  state which is at the potential minima where the probability of downward bound-free transition by spontaneous emission 
 is maximum due to the largest CG co-efficient. 
 The previously gained potential energy is  carried away by the emitted photon. As a consequence, the atom-pair loses its total kinetic energy. Now, if ${\mathbf k}_{\rm c.m.}\simeq0$, most of the energy lost is the relative energy.   
 The c.m. motion now  repeats climbing  up the next hill and so on. 
 In this process, eventually the atom-pairs  become cooled and trapped in the minima of the potential. 
\vspace{.5cm}
 \begin{figure}[h!]
                       \includegraphics[width=0.45\textwidth]{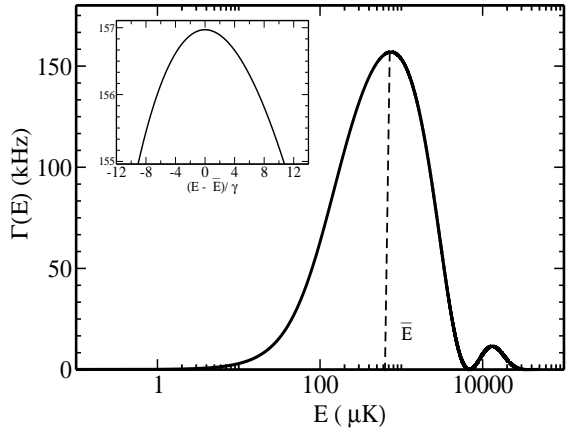}
                       \caption{ $\Gamma(E)$ as a function of  relative kinetic energy $E$.  }  
                       \label{fig:fig2}
            \end{figure}

 \section{Results and discussions} 
 
 Next, we discuss practical implementation of our proposed scheme. A set of good internal states (molecular angular momenta) of two colliding atoms 
represent a scattering channel. It is preferable to have only one ground-state scattering channel that can be subjected to Sisyphus cooling cycle  so that 
atoms have no chance to be transferred to other ground-state channel by spontaneous emission \cite{pra82:2010,pra90:2014}. Alternatively, the energy difference between asymptotic 
thresholds of different ground-state channels should be large enough compared to bound-free spontaneous linewidth so that atoms can be cycled 
only between the chosen (cooling) channel and the excited bound state. Although Sisyphus mechanism will predominantly 
bring the atom-pair into the potential minima reducing the kinetic energy, there is still finite 
probability of transitions to two free atoms into the continuum with higher energies due to spontaneous emission.  
In order to recycle these higher energy atoms back into cooling cycles the linewidth of
PA lasers should be larger than the spontaneous linewidth $\gamma$.

For numerical illustration, we consider fermionic $^{171}$Yb atoms which have  $p$-wave PLR 
  excited states that are recently used to demonstrate $p$-wave optical Feshbach resonance \cite{prl100:2008,pra82:2010,pra90:2014}.  
  For these PLR molecular states,  the  
    projection $\Phi$ of the total angular momentum ${\mathbf F} = {\mathbf J} + {\mathbf I}$ on the internuclear axis is a good quantum number. 
  Further, when one includes the rotation $\ell$ of the relative motion between the two atoms,  
  the  good quantum numbers are  ${\mathbf T} = {\mathbf F}
   + \vec{\ell} $ and its projection $m_T $ on the space-fixed axis.  
   We consider  0$^-$   ($\Phi_e = 0$)  PLR state for which only  odd  $T_e$ are allowed 
   \cite{prl100:2008,pra:1996:Abraham,pra:2005:Tiesinga}. These PLR states are accessible by PA transitions only from odd partial-waves 
    (odd $\ell$) and nuclear spin triplet ($I=1$). 
    Let the two counter-propagating PA lasers be tuned near resonance to $T_g=2 \leftrightarrow T_e=3$ transition
that is dominated by $p$-wave ($\ell=1$) contribution at low energy. 
We choose vibrational quantum number $\nu=1$ for $T_e=3$ having binding energy $-355.4$ MHz 
below the threshold of the corresponding PLR potential.
By selection rules, $T_e=3$ is accessible via PA  only from $T_g=2$ at low energy. Because, from the consideration of fermionic symmetry of the 
two ground-state $^{171}$Yb atoms, $s$- and $d$- and all higher even partial-wave ground-state scattering states will be associated with nuclear spin-triplet 
state ($I=0$). Optical dipole transitions to $T_e=3$ from these two states will be 
forbidden since the molecular electronic wave functions (molecular orbitals) of these two ground states and the excited state belong to the 
same symmetry (even) under reflection at the midpoint of the internuclear axis. On similar arguments, 
spontaneous emission from $T_e = 3$ to all even partial-wave ground-state channels (i.e, $s$-
and $d$-wave)
is forbidden  \cite{prl100:2008}.  The molecular dipole moment 
${\mathbf D}$ will couple ground and excited molecular electronic states of opposite electronic center-of-symmetry. Since the excited electronic state 
we consider here 0$^-$   ($\Phi_e = 0$) has positive symmetry \cite{prl100:2008}, the ground electronic state of negative symmetry only will be coupled to this excited state. The molecular 
ground-state of all even partial waves has  positive symmetry due to Fermionic symmetry of the two nuclei with $I=0$, 
and therefore spontaneous emission from excited state to $s$-, $d$- and higher even partial-wave channels   
are ruled out.  $f$-wave channel is not dominant at low energy and so can be ignored.

The dependence of stimulated linewidth $\Gamma(E)$ on $E$ is shown in 
Fig.2. $\Gamma(E)$ is proportional to the square of FC factor. Figure 2 exhibits that $\Gamma(E)$ attains a prominent maximum near $E \simeq 750 \mu$K with a 
broad width. This feature can be attributed to the nature of  PLR bound state which can be accessible from almost 
asymptotic regime of ground-state continuum. From Fig.2, we notice that at $E = 3$ mK, 
the FC factor reduces to about one tenth of its peak value near  $E = 0.75$ mK. As discussed in the preceding section, successive absorption-spontaneous emission cycles will
on an average reduce the relative kinetic energy. Therefore, the spontaneous emission will not
be a major hindrance to cooling as long as there are no other states except the single-channel continuum
to which the atom-pair can decay. Suppose, initial relative energy is 750 $\mu K$ at which the PA coupling
is maximized as Fig.2. shows. Although the atom-pair will be preferentially transferred towards lower
relative energy side, there will be still some finite probability that spontaneous emission will push two atoms to
higher energy side. To bring them back into cooling cycle, a pair of broadband PA lasers (see the 
Introduction section). Since $\gamma\simeq$ 364 kHz, repumping lasers with bandwidth of a few MHz will suffice the purpose. The insert to Fig.2 shows that the peak structure of PA coupling at $E=\bar{E}$ is sufficiently broad compared to $\gamma$. At energies $E = \bar{E} \pm 10 \gamma$, the stimulated 
PA linewidth (or equivalently square of the FC factor) changes by about 1\% only from its peak value at $E=\bar{E}$. This weak dependence of PA couping on 
$E$ near the resonance energy $\bar{E}$ when the laser is tuned to the resonance frequency $\bar{\omega}_L$ [$=(\bar{E}/\hbar + \omega_b$] facilitates us 
to obtain an approximate analytical solution of the master equation in the steady-state  as discussed in Appendix-B. 
Figure 2 further shows that for $E < 8 \mu K$, PA coupling is vanishingly small. This is due to threshold effect.
In the limit $E \rightarrow 0$, FC factor
goes to zero. As PA coupling goes to zero, cooling will stop.

\begin{figure}
                       \includegraphics[width=0.45\textwidth]{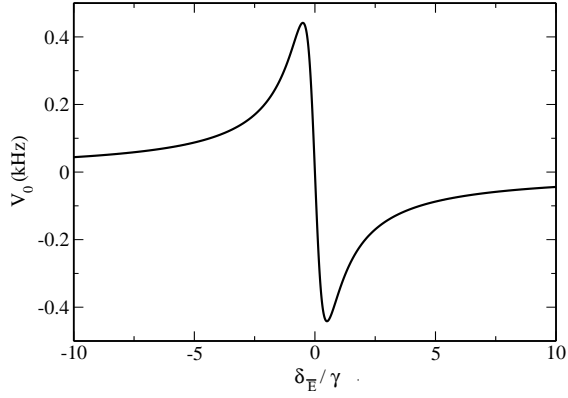}
                       \caption{Strength $V_0$ (in kHz) of the potential $U_{opt}$ is plotted against detuning $\delta_{\bar{E}}$. }  
                       \label{fig:fig3}
            \end{figure}

As a result of photoassociative Sisyphus process, 
we get five spin-dependent c.m. potentials of pair of $^{171}$Yb atoms for ground state $T_g=2$ as given by    
 \begin{eqnarray}
    U_{opt}{(m_{T_g}=\pm2)}&=&-\frac{V_0}{2}[-\frac{8}{7}\pm \cos(2k_L Z)]\nonumber\\
    U_{opt}{(m_{T_g}=\pm1)}&=&-\frac{V_0}{4}[-\frac{13}{7}\pm \cos(2k_L Z)]\nonumber\\
    U_{opt}{(m_{T_g}=0)}&=&-\frac{6}{14}V_0\nonumber\\
    \end{eqnarray}
 where $V_0=-\frac{14}{75}\hbar \delta_{\bar{E}} S_0$. The c.m. motion can be trapped in $m_{T_g}=\pm2,\pm1$ state provided 
 $\delta_{\bar{E}}$ is negative as can be seen from Fig.{\ref{fig:fig3}} which shows that the $U_{opt }$ is attractive (repulsive)
 when $\delta_{\bar{E}} < 0 (>0)$. 
 We take laser intensity 10 mW/cm$^2$. Figure{\ref{fig:fig3}} shows that 
 $V_0$ varies with $ \delta_{\bar{E}} $ reaching a maximum of 
 about 440 Hz at $\delta_{\bar{E}}=0.5 \gamma$, where we have set $\gamma = 364 $ kHz which is taken to be double the 
 atomic linewidth of 182 kHz \cite{pra:2013:yamazaki}. 
 When $\delta_{\bar{E}} $ is negative $V_0$ is positive implying the existence of trapping potential,  
 but when $\delta_{\bar{E}} $ is positive $V_0$ becomes negative showing no trapping is possible. 
 The c.m. recoil limit for$^{171}$Yb$_2$ is $E_{r}^{{\rm c.m.}} = 1.3$ kHz. The  parameter $S_0 = 0.013$ and $S_0 =0.0006 $ 
 for $\delta_{\bar{E}}/\gamma = - 0.5 $ and   - 10 , respectively. In near-resonant 
 case ($\delta_{\bar{E}} \sim \gamma$),  the c.m. motion will be subjected to cooling due to Sisyphus process while for far-off resonant case 
 ($|\delta_{\bar{E}}| >> \gamma$) the c.m. motion will be trapped in shallow potentials. With the given intensity of PA laser we can cool the atom pairs to about a few microkelvin.

 \section{conclusion} 
 
 In conclusion, we have proposed a Sisyphus-like method with photoassociative transitions for optically cooling and trapping an atom-pair by optical dipole force that acts on the c.m. of the pair.
 As in the case of molecules, laser cooling of a pair of interacting atoms is a challenging problem. There is
no general method of laser cooling of molecules due to the existence of a large number of closely lying
ro-vibrational levels. Only in some specific cases of favorable level structure, laser cooling of molecules is possible.
Similarly, to cool an interacting pair of atoms by photoassociative Sisyphus-like laser cooling technique,
one has to look for some favorable atomic systems which have a single asymptotic collision channel with degenerate
molecular magnetic sublevels and narrow line PA transitions. The atoms should be pre-cooled so that the c.m. momentum is small enough
to be subjected to Sisyphus mechanism with counter-propagating PA lasers. 
We have shown
that fermionic $^{171}$Yb is one such favorable PA system due to the existence of narrow-line PA transitions. Our method critically depends
on the strength and nodal structure of the free-bound overlap integral at low energy. In this context,
we have shown PLR states play an important role. We have also discussed repumping  of the atom-pair with relatively high energy 
into the cooling cycle by using broadband PA lasers in $\sigma_+ + \sigma_-$ polarization configuration. The use of spectrally filtered broadband 
lasers \cite{naturecomm:2014,PRA:80:2009:Sofikitis} will be particularly useful for our purpose. For instance, suppose all frequency components higher than the resonant frequency $\bar{\omega}_L$ 
are filtered out from the repumping lasers. Then the repumping lasers will recycle  into the cooling cycles  those atom-pairs which 
have energies higher than the resonant energy $\bar{E}$. In this context, it is to be born in mind that while Sisyphus lasers will optically 
pump the atom-pairs in different ground-state sub-levels, broadband repumping lasers are needed only to recycle the atom-pairs 
which go far-off resonant on the higher side of the energy due to spontaneous emission. Apart from cooling and trapping of atom-pairs, 
our proposed method will be particularly important for preparing  a system of spatially isolated ultracold atom-pairs for conversion into molecules 
 with spatial control over association process. This method will also be applicable for manipulating  interactions between two atoms
 with trapped c.m. motion. Particularly important prospect for our method will be a scenario where one can simultaneously cool and manipulate on-site 
 interactions in a Mott-insulator of atoms in optical lattice, and thereby to explore new aspects of many-body physics with ultracold atoms. 
 By this way, in near future it might be possible to explore $p$-wave or $d$-wave pairing or superfluidity with fermionic $^{171}$Yb or $^{40}$K atoms 
 in an optical lattice. In essence, our proposed scheme and theoretical method will stimulate further studies opening 
 new perspectives in research with cold atoms and molecules. 
 
 \appendix{}

 \section{The c.m. dipole force}

    The equation of motion for c.m. momentum ($P_{\rm c.m.}$) is given by $ dP_{\rm c.m.}/dt=\frac{i}{\hbar}[H_I,P_{\rm c.m.}]$ which yields $ 
     dP_{\rm c.m.}/dt=\nabla_R \left(\int_0^\infty\Lambda_{bE^{'}}(E',R)e^{-i\phi(r,R)} e^{-i\omega_{L} t}|b\rangle \langle E^{'}|dE^{'}+H.c \right)$. As a result, the c.m. of the two atoms experiences an optical force $ F=\Big\langle dP_{\rm c.m.}/dt\Big\rangle$ , where $\langle\cdots\rangle$ implies averaging over initial states. The c.m. and relative motions of the system of two atoms become coupled due to photoassociative coupling. We are dealing with cold atoms for which the time scale of evolution of c.m. motion can be assumed to be much less than that of relative motion. We can then safely decouple the two degrees of motions and obtain 
 \begin{eqnarray}
     {\mathbf F} &=&\int_0^\infty{\mathbf {\nabla}}_R \Big\{\Big\langle|b\rangle \langle E^{'}|\Big\rangle_{rel}\Big\langle \left(\Lambda_{bE^{'}}(R)e^{-i\phi(r,R)}\right) e^{-i\omega_{L} t}\Big\rangle_{\rm c.m.} dE^{'}+c.c.\Big\}
    \end{eqnarray} 
    One can relate $ \Big{\langle}|b\rangle \langle E^{'}|\Big{\rangle}_{rel}$ to density matrix element $\rho_{E'b}$ as derived in Appendix B.  So, the optically generated force is given by Eq. (\ref{for})
     It is convenient to express Eq.(\ref{for}) in a frame rotating at the frequency $\omega_L $ of the laser field. This is achieved by introducing the  variable $ \rho_{Eb}= {\rho}_{Eb} e^{i\omega_Lt}$. We can thus write $F=F_{dip}+F_{dis}$, where
     
    \begin{eqnarray}
    {\mathbf F}_{dip}&=&\int_0^\infty \Big[e^{-i\phi(r,R)}{\mathbf {\nabla}}_R\Big\{{\Lambda_{bE'}}(R) {\rho}_{E'b} \Big\}+e^{i\phi(r,R)}\nabla_R\Big\{{\Lambda_{E'b}{\rho}_{bE'}}(R) \Big\}\Big]dE' \label{f}\\    
    {\mathbf F}_{dis} &=& \int_0^\infty i\Big\{{\mathbf {\nabla}}_R\phi(r,R)\Big\} \Big[ {\rho}_{bE'} e^{i\phi(r,R)}{\Lambda_{E'b}}(R)  -{\rho}_{E'b}e^{-i\phi(r,R)}{\Lambda_{bE'}}(R)  \Big]dE'   
    \end{eqnarray}
    For plane wave only dissipative force and for standing wave  only dipole force exists. In the later case $\phi(r,R)=0$. We are interested in conservative dipole force,  hence we consider standing wave ($\phi(r,R)=0$).
    
    In order to calculate the forces, we need to evaluate the free-bound density matrix element  $\rho_{E b}$. With relaxation processes  
    included, the atomic density matrix is given by the solution of the master equation for the  density matrix. For an all discreet atomic system, 
    the master equation is well-known. For a coupled discrete-continuum system such as in the present case, density matrix approach 
    is not adequately addressed in the literature.

    Let $\rho_c=\hbar \int \rho_{EE}d\omega_E$  \hspace*{0.1in}    and $\rho_{bc}={\int \limits_{0}^{\infty}} \Lambda_{Eb}(R){\rho}_{bE}d\omega_E$. 
   Using the normalization condition $\rho_{bb}+\rho_c=1$,  we can write 
   \begin{eqnarray}    
          \dot {\rho}_{c}&=&i\rho_{bc}-i\rho_{cb}-\gamma\rho_{bb}\\ 
         \end{eqnarray}  
   Now we get from density matrix equation (appendix B ) 
    \begin{eqnarray} 
    \dot{\rho}_{bc}= \Big(i\Delta-\frac{\gamma}{2}\Big){\rho}_{bc}+ i\hbar {\int \limits_{0}^{\infty}}\Lambda_{Eb}(R)\rho_{bE}\omega_E d\omega_E+{\int \limits_{0}^{\infty}}\Lambda_{bE}(R)\Big\{{\int \limits_{0}^{\infty}}\Lambda_{bE'}(R)\rho_{E'E}d\omega_{E'}\Big\}d\omega_E-\frac{i}{2\pi}{\int \limits_{0}^{\infty}}{\cal G}_{E}\rho_{bb}d\omega_E\nonumber\\ 
    \label{rhobc}
    \end{eqnarray}
    
   The solution of density matrix ( appendix B)  equation of $ \dot{\rho}_{E'E}(t)$ can be formally written as
    \begin{eqnarray}
      \rho_{E'E}(t)&=& {\rho}_{E'E}(0)e^{i\omega_{EE'}t}+\frac{i}{\hbar}e^{i\omega_{EE'}t}\Big[\Lambda_{E'b}{\int \limits_{0}^{t}}{\rho}_{bE}e^{i\omega_{E'E}t'}dt'-\Lambda_{bE}{\int \limits_{0}^{t}}{\rho}_{E'b}(R)e^{i\omega_{E'E}t}dt'\Big]\nonumber\\
      &&+\gamma_{0}\eta_{bE'}^*\eta_{bE}{\int \limits_{0}^{t}}\rho_{bb}(t')e^{i\omega_{E'E}(t'-t)}dt'
    \end{eqnarray}
    where $\omega_{EE'}=\frac{E-E'}{\hbar}$.\\
    Putting the expression $\rho_{E'E}(t)$ in Eq.(\ref{rhobc})
    \begin{eqnarray} 
    \dot{\rho}_{bc}&=& \Big(i\Delta-\frac{\gamma}{2}\Big){\rho}_{bc}+ i\hbar {\int \limits_{0}^{\infty}}\Lambda_{Eb}(R)\rho_{bE}\omega_E d\omega_E-\frac{i}{2\pi}{\int \limits_{0}^{\infty}}{\cal G}_{E}\rho_{bb}d\omega_E\nonumber\\
    &&+i{\int \limits_{0}^{\infty}}\Lambda_{Eb}(R)\Big[{\int \limits_{0}^{\infty}}\Lambda_{bE'}(R)\Big(\rho_{E'E}(0)e^{i\omega_{EE'}t}+\gamma_{0}\eta_{bE'}^*\eta_{bE}{\int \limits_{0}^{t}}\rho_{bb}(t')e^{i\omega_{E'E}(t-t')}dt'\Big)d\omega_{E'}\Big]d\omega_E\nonumber\\
    &&+\frac{1}{\hbar}{\int \limits_{0}^{t}}\Big[{\int \limits_{0}^{\infty}}|\Lambda_{Eb}|^2\Big({\int \limits_{0}^{\infty}}\Lambda_{bE'}\rho_{E'b}e^{i\omega_{E'E}(t-t')}d\omega_{E'}\Big)d\omega_E\Big]dt'\nonumber\\
&&-\frac{1}{\hbar}{\int \limits_{0}^{t}}\Big[{\int \limits_{0}^{\infty}}\Lambda_{Eb}\rho_{bE}\Big({\int \limits_{0}^{\infty}}|\Lambda_{E'b}|^2e^{i\omega_{E'E}(t-t')}d\omega_{E'}\Big)d\omega_E\Big]dt'
    \end{eqnarray}
      
 We assume $\rho_{E'E}(0)=\rho_c(0) \delta(E'-E)$. While doing the energy integrals,  we have assumed that $\delta_E\simeq\delta_{\bar{E}}$. 
This approximation can be justified in the following way.  Since the coherence term $\rho_{b E}$ has the line width $\gamma$ which is quite small compared to the scale of energy variation of PA coupling 
 $\Lambda(E)$ near the resonant energy $\bar{E}$ as discussed in the theory section (see Fig.2), $\rho_{b E}$ can be taken as rapidly varying 
 function of energy compared to all other energy-dependent quantities:

 \begin{eqnarray}
  \dot{\rho}_{bc}&=& 
   \Big(i\delta_{\bar{E}}-\frac{\gamma}{2}\Big){\rho}_{bc} -\frac{i}{2\pi}{\int \limits_{0}^{\infty}}{\cal G}_{E}(R)d\omega_E\Big[2\rho_{bb}-1\Big]
 \end{eqnarray}

 After doing some algebra, we get the steady state solutions
 \begin{eqnarray}
  \rho_{bb}=\frac{\frac{1}{2\pi}{\int \limits_{0}^{\infty}}{\cal G}_{E}(R)d\omega_E}{\delta_{\bar{E}}^2+\frac{\gamma^2}{4}+\frac{2}{2\pi}{\int \limits_{0}^{\infty}}{\cal G}_{E}(R)d\omega_E} \label{rhobbf}\\
  \rho_{bc}=-\frac{\frac{1}{2\pi}{\int \limits_{0}^{\infty}}{\cal G}_{E}(R)d\omega_E\Big(\delta_{\bar{E}}-i\frac{\gamma}{2}\Big)}{\delta_{\bar{E}}^2+\frac{\gamma^2}{4}+\frac{2}{2\pi}{\int \limits_{0}^{\infty}}{\cal G}_{E}(R)d\omega_E}
 \end{eqnarray}
 From the dipole force from Eq. (\ref{f}) we get
\begin{eqnarray}
 F_{dip}
 &=&-\frac{\hbar\delta_{\bar{E}}}{\pi}\Big[\frac{\nabla_R{\int \limits_{0}^{\infty}}{\cal G}_{E}(R)d\omega_E}{\delta_{\bar{E}}^2+\frac{\gamma^2}{4}+\frac{2}{2\pi}{\int \limits_{0}^{\infty}}{\cal G}_{E}(R)d\omega_E}-\frac{\frac{2\hbar}{2\pi}{\int \limits_{0}^{\infty}}{\cal G}_{E}(R)d\omega_E(\nabla_R{\int \limits_{0}^{\infty}}{\cal G}_{E'}(R)d\omega_{E'})}{\Big(\delta_{\bar{E}}^2+\frac{\gamma^2}{4}+\frac{2}{2\pi}{\int \limits_{0}^{\infty}}{\cal G}_{E}(R)d\omega_E\Big)^2}\Big]
 \label{fdip}
\end{eqnarray}
 where ${\cal G}_E(R)= \frac{2\pi}{\hbar}|\Lambda_{bE}(R)|^2 =\Gamma(E)\Theta(R)$.\\
 We define the saturation parameter $ S_0=\frac{\frac{1}{\pi}{\int \limits_{0}^{\infty}}{\Gamma}(E)d\omega_E}{\delta_{\bar{E}}^2+\frac{\gamma^2}{4}}$. At the weak intensity limit the second term on the right side $F_{dip}$ Eq. (\ref{fdip}) can be neglected. At weak intensity$(i.e   S_0<<1)$ the dipole force becomes 
 \begin{eqnarray}
   F_{dip}(R)=-\frac{\hbar \delta_{\bar{E}}}{\pi}\Big[\frac{\nabla_R{\int \limits_{0}^{\infty}}{\cal G}_{E}(R)d\omega_E}{\delta_{\bar{E}}^2+\frac{\gamma^2}{4}+\frac{2}{2\pi}{\int \limits_{0}^{\infty}}{\cal G}_{E}(R)d\omega_E}\Big]
 \end{eqnarray}

The optical potential is
\begin{eqnarray}
 U_{opt}&=&-\int F_{dip}(R).dR\nonumber\\
 &=&\hbar\delta_{\bar{E}}\ln \Big(1+\frac{\frac{1}{\pi}{\int \limits_{0}^{\infty}}{\cal G}_{E}(R)d\omega_E}{\delta_{\bar{E}}^2+\frac{\gamma^2}{4}}\Big)
\end{eqnarray}
 At weak intensity$(S_0<<1)$ The  optical potential is
 \begin{eqnarray}
  U_{opt}=\hbar\delta_{\bar{E}}\frac{\frac{1}{\pi}{\int \limits_{0}^{\infty}}{\cal G}_{E}(R)d\omega_E}{\delta_{\bar{E}}^2+\frac{\gamma^2}{4}}
 \end{eqnarray}

In terms of $S_0$ Eq. (\ref{rhobbf}) can be written as
\begin{eqnarray}
 \rho_{bb}=\frac {S_0 \Theta(R)}{2+2S_0 \Theta(R)}
\end{eqnarray}
When $S_0\rightarrow \infty$, $\rho_{bb}\rightarrow \frac{1}{2}$  
    
  \section{Derivation of the master equation for PA system}
  
  To incorporate spontaneous decay in the continuum-bound system under consideration, 
  we derive the master equation of the the system from  first principles.  
  The excited molecular bound state can spontaneously decay to the continuum scattering states of two ground-state atoms 
  due to coupling of the system with a reservoir of electromagnetic vacuum modes.
  The Hamiltonian is $H = H_{S} + H_R + H_{SR}$, where  $H_S \equiv\hat{H_0}^{{\rm rel}}$, the system Hamiltonian as given in Eq.(\ref{hrel}).
  $H_R = \sum_{\kappa\sigma} \hbar \omega_{\kappa\sigma} \hat{a}_{\kappa\sigma}^{\dagger} \hat{a}_{\kappa\sigma}  $ represents the Hamiltonian 
  for the reservoir of the infinite number of electromagnetic modes denoted by the wave number $\kappa$ and the polarization $\sigma$ 
  with $ \hat{a}_{\kappa\sigma} (\hat{a}_{\kappa\sigma}^{\dagger}) $ denoting photon annihilation (creation) operator 
  for the (${\kappa\sigma}$) mode. Here we are considering the incoherent interaction of the vacuum modes only. We have dropped PA laser interaction part of the Hamiltonian for the sake of simplicity. The system-reservoir interaction Hamiltonian can be expressed as 
  \bea 
H_{SR} &=& \left [ \sum_{\kappa\sigma} \hat{a}_{\kappa\sigma}e^{-i\omega_{\kappa\sigma}t}  
 \int d E {\cal V}_{E}(\kappa\sigma) 
   S_E^+ e^{i\omega_{bE}}    +  \rm{H.c.} \right ] 
   \label{hsr}
\eea 
 where $\omega_{bE}=E/\hbar+\omega_{b}$, with $\omega_{b}$ being the frequency of the bound state measured from the threshold of the continuum. ${\cal V}_E(\kappa\sigma)=-E_{vac}(\kappa)(\vec D.\hat {\epsilon}_\sigma)\eta_{bE}$ is the vacuum-induced continuum-bound coupling with $E_{vac}(\kappa)=\sqrt{\frac{\hbar\omega_{\kappa}}{2\epsilon_0 V}}$, $\hat{\epsilon_{\sigma}}$ being the unit vector of polarisation and V is the volume, $\vec D$ is the molecular Dipole moment operator and $\eta_{bE}=\int_0^{\infty}d^3r\phi_b(\vec r)\psi_E(\vec r)$ is the bound-free overlap integral, $\phi_b(\vec r)$ and $\psi_E(\vec r)$ being respectively the bound and free wave function.
 
 The equation for the incoherent part of the reduced density matrix of the system \cite{book:carmichhal} is
 \bea
 \dot{\rho}_{inc}=-\int_0^t dt' {\text Tr}_R \left \{[H_{SR}(t),[H_{SR}(t'),\rho(t')R_0]]     \right\}
   \label{rhodotinc}
 \eea
 $R_0$ is an initial reservoir density operator. Substituting Eq.(\ref{hsr}) in Eq.(\ref{rhodotinc}), and expanding the commutator, we have

 \begin{eqnarray}
  \dot{\rho}_{inc}&&=-\sum_{\kappa\sigma\kappa '\sigma '}\int_0^{\infty}dE \int_0^{\infty}dE'\int_0^t dt'         {\text Tr}_R { \Bigg\{} {\Bigg(}{\Big[}{\cal V}^*_E(\kappa\sigma){\cal V}^*_{E'}(\kappa '\sigma ')S_ES_{E'} \rho(t')R_0\nonumber\\
  &&- {\cal V}^*_E(\kappa\sigma){\cal V}^*_{E'}(\kappa '\sigma ')S_E\rho(t')S_{E'} R_0 {\Big]}e^{-i(\omega_{bE}t+\omega_{bE'}t')}\hat{a}^+_{\kappa\sigma}\hat{a}^+_{\kappa '\sigma '}e^{i(\omega_{\kappa\sigma}t+\omega_{\kappa'\sigma'}t')}       +{\rm H.c.}{\Bigg)}\nonumber\\
 &&+{\Bigg(}{\Big[}{\cal V}_E(\kappa\sigma){\cal V}_{E'}(\kappa '\sigma ')S^+_ES^+_{E'} \rho(t')R_0
 - {\cal V}_E(\kappa\sigma){\cal V}_{E'}(\kappa '\sigma ')S^+_E\rho(t')S^+_{E'} R_0 {\Big]}\nonumber\\
 &&\times e^{i(\omega_{bE}t+\omega_{bE'}t')}\hat{a}_{\kappa\sigma}\hat{a}_{\kappa '\sigma '}e^{-i(\omega_{\kappa\sigma}t+\omega_{\kappa'\sigma'}t')}       +{\rm H.c.}{\Bigg)}+{\Bigg (}{\Big[}{\cal V}_E(\kappa\sigma){\cal V}^*_{E'}(\kappa '\sigma ')S_ES^+_{E'} \rho(t')R_0\nonumber\\
 &&- {\cal V}^*_E(\kappa\sigma){\cal V}_{E'}(\kappa '\sigma ')S^+_E\rho(t')S_{E'} R_0 {\Big]}e^{-i(\omega_{bE}t-\omega_{bE'}t')}\hat{a}^+_{\kappa\sigma}\hat{a}_{\kappa '\sigma '}e^{-i(\omega_{\kappa\sigma}t-\omega_{\kappa'\sigma'}t')}       +{\rm H.c.}{\Bigg)}\nonumber\\
 &&{\Bigg(}{\Big[}{\cal V}^*_E(\kappa\sigma){\cal V}_{E'}(\kappa '\sigma ')S^+_ES_{E'} \rho(t')R_0- {\cal V}_E(\kappa\sigma){\cal V}^*_{E'}(\kappa '\sigma ')S_E\rho(t')S^+_{E'} R_0 {\Big]}\nonumber\\
 && \times e^{i(\omega_{bE}t-\omega_{bE'}t')}\hat{a}_{\kappa\sigma}\hat{a}^+_{\kappa '\sigma '}e^{-i(\omega_{\kappa\sigma}t-\omega_{\kappa'\sigma'}t')}       +{\rm H.c.}{\Bigg)}{ \Bigg\}}
 \label{rhoinc}
   \end{eqnarray}
   Tracing over the vacuum modes \cite{springer74gsa} and using 
   
\begin{eqnarray}
&&{\text Tr}\{R_0 \hat{a}_{\kappa\sigma}\hat{a}^+_{\kappa '\sigma '}\}=\delta_{\kappa\kappa '}\delta_{\sigma\sigma '},\hspace*{2cm} {\text Tr}\{R_0 \hat{a}^+_{\kappa\sigma}\hat{a}_{\kappa '\sigma '}\}=0\nonumber\\
&&{\text Tr}\{R_0 \hat{a}_{\kappa\sigma}\hat{a}_{\kappa '\sigma '}\}={\text Tr}\{R_0 \hat{a}^+_{\kappa\sigma}\hat{a}^+_{\kappa '\sigma '}\}=0
   \end{eqnarray}
we are left with the last term containing $\hat{a}_{\kappa\sigma}\hat{a}_{\kappa '\sigma '}^+$. After making the change of variable $\tau=t-t'$ Eq. (\ref{rhoinc}) reduces to 
 \begin{eqnarray}
  \dot{\rho}_{inc}&&=-\sum_{\kappa\sigma}\int_0^{\infty}dE \int_0^{\infty}dE'\int_0^t d\tau {\Big\{}{\cal V}^*_E(\kappa\sigma){\cal V}_{E'}(\kappa\sigma)S^+_ES_{E'} \rho(t-\tau)\nonumber\\&& - {\cal V}_E(\kappa\sigma){\cal V}^*_{E'}(\kappa\sigma)S_E\rho(t-\tau)S^+_{E'}{\Big\}}
 e^{i\omega_{EE'}t}e^{i\omega_{bE'}\tau}e^{-i\omega_{\kappa\sigma}\tau}       +H.c.
   \end{eqnarray}
Under Markov approximation we get 
 \begin{eqnarray}
  \dot{\rho}_{inc}&&=-\sum_{\kappa\sigma}\int_0^{\infty}dE \int_0^{\infty}dE'e^{i\omega_{EE'}t}{\Big\{}S^+_ES_{E'} \rho(t){\cal V}^*_E(\kappa\sigma){\cal V}_{E'}(\kappa\sigma)\int_0^t d\tau e^{i(\omega_{bE'}-\omega_{\kappa\sigma})\tau}\nonumber\\&& - {\cal V}_E(\kappa\sigma){\cal V}^*_{E'}(\kappa\sigma)S_E\rho(t)S^+_{E'}\int_0^t d\tau e^{i(\omega_{bE'}-\omega_{\kappa\sigma})\tau}{\Big \}}
        +H.c.
   \label{rho}
   \end{eqnarray}
   We can write ${\cal V}_{E'}(\kappa\sigma)=g_{\kappa\sigma}\eta_{bE}$ where $g_{\kappa\sigma}=-E_{vac}(\kappa)(\vec D.\hat {\epsilon}_\sigma)$ and ${\cal V}_E(\kappa\sigma){\cal V}^*_{E'}(\kappa\sigma)=|g_{\kappa\sigma}|^2\eta_{bE}\eta^*_{bE'}$. Now we convert the sum $\sum_{\kappa\sigma}$ into an energy integral using the relation 
   \begin{eqnarray}
   \sum_{\kappa\sigma} |g_{\kappa\sigma}|^2= \int \frac{D^2}{6\pi^2\epsilon_0\hbar c^3}\omega_{\kappa}^3d\omega_{\kappa}
 \end{eqnarray}
 For bound-continuum system, the spontaneous linewidth can be defined as $\gamma=\gamma_0\int_0^{\infty}dE|\eta_{bE}|^2$, where $\gamma_0=\frac{D^2}{3\pi\epsilon_0\hbar c^3}\omega_b^3$.
 
  As the time of interest $t\gg 1/\omega_{bE}$, where $\omega_{bE}$ is generally in the optical frequency domain we can take upper limit of above time integral to $\infty$. Writing the integral as 
      \begin{eqnarray}
    \int_0^\infty d\tau e^{i(\omega_{bE^{'}}-\omega_{\kappa\sigma}) \tau}&=&\lim_{\epsilon\rightarrow0}  \int_0^\infty d\tau e^{i(\omega_{bE^{'}}-\omega_{\kappa\sigma}+i\epsilon) \tau}\nonumber\\
    &=&-\lim_{\epsilon\rightarrow0}\frac{1}{i\left(\omega_{bE^{'}}-\omega_{\kappa\sigma}+i\epsilon\right)}\nonumber\\
    &=&  \pi \delta \left(\omega_{\kappa\sigma}-\omega_{bE^{'}}\right)+i{\cal P}\left(\frac{1}{\omega_{\kappa\sigma}-\omega_{bE^{'}}}\right)
    \end{eqnarray}
    where ${\cal P}$ represents the Cauchy principal part which leads to Lamb shift. This is quite small and so can be ignored for the present purpose. So we write Eq. (\ref{rho}) as
    \begin{eqnarray}
      \dot{\rho}_{inc}&&=-\int_0^{\infty}dE \int_0^{\infty}dE'e^{i\omega_{EE'}t}\frac{\gamma_0}{2}{\Big\{}S^+_ES_{E'} \rho(t)\eta_{bE}^*\eta_{bE'}-S_E\rho(t)S^+_{E'}\eta_{bE'}^*\eta_{bE}{\Big \}}\nonumber\\
      &&   -\int_0^{\infty}dE \int_0^{\infty}dE'e^{-i\omega_{EE'}t}\frac{\gamma_0}{2}{\Big\{} \rho(t)S^+_{E'}S_E\eta_{bE'}^*\eta_{bE}-S_{E'}\rho(t)S^+_E\eta_{bE}^*\eta_{bE'}{\Big \}}
      \label{rho1}
    \end{eqnarray}
    Using Eq. (\ref{rho1}) we get
  \begin{eqnarray}
      ({\dot\rho}_{bb})_{inc}&&=-\gamma\rho_{bb}\nonumber\\
       ({\dot\rho}_{bE})_{inc}&&=-\frac{\gamma}{2}\rho_{bE}\nonumber\\
       ({\dot\rho}_{EE'})_{inc}&&=\gamma_{0}\rho_{bb}\eta_{bE'}^*\eta_{bE}\nonumber\\
    \end{eqnarray}

 The complete Master equation for our system is ${\dot\rho}={\dot\rho}_{coh}+{\dot\rho}_{inc}$, where
  \begin{eqnarray}
      {\dot\rho}_{coh}=-\frac{i}{\hbar}[H_I^{coh},\rho]
     \end{eqnarray}
 with $H_I^{coh}=\left[\int_0^\infty\Lambda_{bE^{'}} |b\rangle \langle E^{'}|dE^{'}+H.c\right]$ is the interaction Hamiltonian in the interaction picture due to PA laser only. Thus, for continuum-bound coupled system as in the present case the density matrix elements are given by
    \begin{eqnarray}
     \hbar \dot{{\rho}}_{bE}&=&\left(i\hbar \delta_E-\frac{\hbar\gamma}{2}\right){\rho}_{bE}+i\left[\int_0^\infty\Lambda_{bE^{'}}(R)
    \hbar \rho_{{E^{'}}E}dE'-\Lambda_{bE}(R)\rho_{bb}\right] \\
      \hbar\dot{\rho}_{E^{'}E}&=&-i(E^{'}-E)\rho_{E^{'}E}+i\Big[\Lambda_{E'b}(R){\rho}_{bE}-{\rho}_{E'b}\Lambda_{bE}(R)\Big]  +\gamma_{0}\rho_{bb}\eta_{bE'}^*\eta_{bE}\\
      \hbar\dot{\rho}_{bb}&=&\left(i\int_0^\infty{\rho}_{E^{'}b}\Lambda_{bE^{'}}(R)dE{'}+c.c.\right)-\hbar\gamma\rho_{bb} 
      \label{mo1}
    \end{eqnarray}
    where $\delta_E=\Delta+\omega_E$ and $\Delta=\omega_L-\omega_0$ ($\omega_L , \omega_0$) is laser  and atomic frequency respectively $\omega_E=E/\hbar$).

 \begin{acknowledgments}
 One of us (BD) is thankful to Olivier Dulieu for helpful discussion and a suggestion. 
 \end{acknowledgments}

\end{document}